\documentclass[aps,pre,reprint,superscriptaddress]{revtex4-2}
\usepackage{natbib}
\usepackage{graphicx,amsmath, xcolor,amssymb,listings,verbatim, hyperref}
\usepackage[utf8]{inputenc}
\usepackage[T1]{fontenc}

\hypersetup{
    colorlinks = false,
    linkbordercolor = {white},
}
\newcommand{\changes}[1]{#1}

\begin{document}

\title{Generating consensus and dissent on massive discussion platforms\\ with a \boldmath semantic-vector model}

\author{Alfredo~Ferrer}
\affiliation{Instituto de Biocomputación y Física de Sistemas Complejos (BIFI), Universidad de Zaragoza, 50018 Zaragoza, Spain}
\affiliation{Kampal Data Solutions,  WTCZ, Avda. Maria Zambrano 31, 50018 Zaragoza, Spain}
\author{David~Muñoz-Jordán}
\email{Contact author: dmunoz@bifi.es}
\affiliation{Instituto de Biocomputación y Física de Sistemas Complejos (BIFI), Universidad de Zaragoza, 50018 Zaragoza, Spain}
\author{Alejandro~Rivero}
\affiliation{Instituto de Biocomputación y Física de Sistemas Complejos (BIFI), Universidad de Zaragoza, 50018 Zaragoza, Spain}
\affiliation{Kampal Data Solutions,  WTCZ, Avda. Maria Zambrano 31, 50018 Zaragoza, Spain}
\author{Alfonso~Tarancón}
\affiliation{Departamento de Física Teórica, Universidad de Zaragoza, 50009 Zaragoza, Spain}
\author{Carlos~Tarancón}
\affiliation{Kampal Data Solutions,  WTCZ, Avda. Maria Zambrano 31, 50018 Zaragoza, Spain}
\author{David~Yllanes}
\affiliation{Instituto de Biocomputación y Física de Sistemas Complejos (BIFI), Universidad de Zaragoza, 50018 Zaragoza, Spain}
\affiliation{Departamento de Física Teórica, Universidad de Zaragoza, 50009 Zaragoza, Spain}
\affiliation{Zaragoza Scientific Center for Advanced Modeling (ZCAM), 50018 Zaragoza, Spain}
\affiliation{Fundación ARAID, Diputación General de Aragón, 50018 Zaragoza, Spain}

\begin{abstract}

Reaching consensus on massive discussion networks is critical for reducing noise and achieving optimal collective outcomes. However, the natural tendency of humans to preserve their initial ideas constrains the emergence of global solutions. To address this, Collective Intelligence (CI) platforms facilitate the discovery of globally superior solutions. We introduce a dynamical system based on the standard $O(N)$ model to drive the aggregation of semantically similar ideas. The system consists of users represented as nodes in a $d=2$ lattice with nearest-neighbor interactions, where their ideas are represented by semantic vectors computed with a pretrained embedding model. 
We analyze the system's equilibrium states as a function of the coupling parameter $\beta$. Our results show that $\beta > 0$ drives the system toward a ferromagnetic-like phase (global consensus), while $\beta < 0$ induces an antiferromagnetic-like state (maximum dissent), where users maximize semantic distance from their neighbors. This framework offers a controllable method for managing the tradeoff between cohesion and diversity in CI platforms.
\end{abstract}

\maketitle
\section{Introduction}
The dynamics of opinion formation in large-scale multi-agent systems is a central theme in sociophysics. In social networks and collaborative environments, ideas compete for dominance, exhibiting evolutionary patterns analogous to physical processes such as percolation \citep{perc:17,jusup:21,bollobas:06}. When bot or hybrid populations are considered, consensus in multi-agent systems emerges as a particularly challenging problem in the field of artificial intelligence~\cite{amirkhani:22}.

The dynamics governing consensus generation have been extensively characterized in statistical physics, where the study of active matter~\cite{marchetti:13,vansaarlos:24} has been particularly fruitful in models and experiments. Indeed, the study of flocks is the foundation
of this field, beginning with the celebrated Vicsek model~\cite{vicsek:95} and developing 
with studies of birds~\cite{ballerini:08, cavagna:14}, bacterial colonies~\cite{zhang:10}
or sheep~\cite{ginelli:15}, to give just a few examples. A vast literature has hence developed on the topics of mutual and self alignment in active matter~\cite{toner:05, julicher:07,vicsek:12,doostmohammadi:18, baconnier:25}. Beyond alignment interactions, the ideas of ``quorum sensing''~\cite{duan:23, lefranc:25} and the motility-induced phase transition~\cite{cates:15} have been important to understand collective phenomena in microbiological systems~\cite{miller:01,liu:19,zhao:23}. The active-matter approach has also found application to human systems, from the interactions of pedestrians~\cite{karamouzas:14} to the dynamics of mosh pits in heavy-metal concerts~\cite{silverberg:13}.

A usual characteristic of active and other non-equilibrium systems
is the lack of reciprocity~\cite{fruchart:21,you:20,ivlev:15,martin:25}, 
which becomes particularly important when considering social groups. As a 
first toy ---but seminal--- example, the Kuramoto model of coupled oscillators can be adapted to 
represent a combination of ``conformists'' and ``contrarians''~\cite{hong:11}. 
More generally, heterogeneity in populations has been shown to have dramatic
effects on collective motion. For instance, in collective migration ``small 
proportions of individuals actively acquir[e] directional information from 
their environment, whereas the majorities use a socially facilitated movement 
behavior''~\cite{guttal:10}. The effect of such 
``opinion leaders'' can be enhanced ---and consensus promoted--- by the 
existence 
of uninformed individuals~\cite{couzin:11} while, on the other hand, a small number
of active dissenters can disrupt alignment interactions and destroy flocks~\cite{yllanes:17}.

\changes{In social physics, opinion dynamics have increasingly focused on non-equilibrium frameworks to explain the emergence of polarization and echo chambers~\cite{Baumann:20, Liu:23}. These models typically characterize an agent's opinion as a unidimensional scalar value where the sign denotes a binary stance—positive or negative—on a specific controversial issue, and the magnitude represents the strength of that conviction. Within these systems, non-equilibrium phase transitions between consensus and polarization are driven by mechanisms such as reinforcement-based dynamics or coevolving networks, where agents preferentially interact with those holding similar opinions.}

\changes{In the contexts discussed thus far, ``consensus'' has traditionally been defined as the alignment of these scalar stances or the convergence of low-dimensional collective behaviors. Recent research has generalized these concepts to multidimensional topic spaces~\cite{Baumann:21, Piao:23}. In these generalized frameworks, an agent's opinion is represented as a vector where each component corresponds to a distinct, though possibly overlapping, theme or topic. These models investigate how feedback-driven polarization or ``information cocoons'' arise from the dynamic adaptation of user preferences across multiple themes.}

\changes{Our approach, however, interprets consensus more literally as the adoption of complex, multifaceted ideas within a social group, rather than a position on a single scale. Unlike models centered on extreme polarization (\emph{e.g.}, pro-choice vs. pro-life), our framework utilizes high-dimensional semantic vectors to represent open-ended responses to a single question. In this semantic space, lack of similarity between two responses does not inherently imply ideological opposition or bipolarization, but simply a lack of semantical agreement.}

Early experiments in small face-to-face groups demonstrated 
the emergence of Collective Intelligence (CI) \citep{woolley:20}, where group interactions lead 
to ideas superior to those of isolated individuals. The digital era has scaled these dynamics,
enabling massive online platforms designed to foster constructive collaboration. Recent 
experiments involving up to a thousand participants have evinced the emergence of solutions 
superior to individual proposals. These benefits span from sensitive social issues, such as 
cyberbullying and gender equality, to deterministic academic problems in mathematics or physics
\citep{orejudo:22,cebollero:22}. 

In these experimental setups, users are typically arranged on a virtual two-dimensional square
lattice with periodic boundary conditions. Each user is presented with a problem and asked to 
propose a solution in the form of a typed string of text. Periodically, participants are allowed
to interact with their four nearest neighbors. Convergence towards a unique global solution 
relies on an imitation mechanism, where superior solutions ideally  propagate through the 
lattice, increasing their domain size over time via successive copying events. At variance with
the directions of motion in a flock, however, it is not immediately 
obvious how to quantify the similarity of two competing ideas or how to propose a dynamical law
that would drive the system towards consensus (or dissent).

Humans and bots exhibit significant inertia. They frequently retain their initial opinions 
rather than adopting their neighbors' views, even when presented with marginally superior
alternatives. To overcome this resistance and facilitate consensus, artificial dynamics are 
often introduced, such as forced copying events, topological rewiring or the extinction of low-frequency responses \citep{garcia-egea:24}. All these rules inherently require a selection bias
favoring certain solutions over others. Defining an objective fitness function is problematic 
for social or philosophical questions, where multiple valid viewpoints coexist and no objective 
``ground truth'' exists. Imposing an external value judgment compromises the system's 
independence as there are no  ``sacred'' solutions. 

In an effort to avoid such value judgments, \citep{garcia-egea:24} proposed 
using the number of times a response appears in the system ---that is, its frequency--- as
a simple goodness criterion. Although this metric proved effective, it presents two 
major limitations. First, it depends on exact string matching, treating semantically 
equivalent but textually distinct responses as unrelated states. Second, it introduces nonlocal effects. A change in the state of a single node instantly alters the global frequency 
distribution, thereby influencing the transition probabilities of all nodes in the system 
simultaneously.

In this work, building on the tools of statistical mechanics and on recent advances
in large language models (LLMs), we propose an approach to overcome these limitations. Language models based on neural networks have undergone a sequence of revolutions, first by increasing 
the hidden size of each layer and the number of layers, and then by expanding the number of
simultaneous tokens that are attended to in each step, which resulted in the
transformers architecture~\citep{attention:17}. Generative models such as GPT specialize 
this architecture to predict the next token, a simple process that allows training hundreds of
billions of parameters. BERT models~\citep{roberta:19} specialize in predicting intermediate deleted tokens, requiring
more complex training, limited to hundreds of millions of parameters, but allowing each token to be aware of the
meaning of all entire sentence. It is then possible to use a large BERT model to extract the semantic
meaning by averaging the hidden sizes across all the tokens of the sentence. Current literature calls the vector
produced in this way an \emph{embedding}, but we prefer to use the classical denomination
and say that each idea will be mapped to a (high-dimensional) \emph{semantic vector}. The BERT model is fine
tuned so that the similarity between two responses 
can be quantified  by the scalar product of their vectors. 

This strategy allows us to 
transition from a frequency-based global coupling to a semantics-based local interaction. From
this local interaction we shall define a total energy for the system ---complete consensus
corresponding to total alignment of the semantic vectors and, hence, to the energy ground 
state. This energy function will allow us to define a (stochastic) dynamical law, where a 
fictitious temperature will regulate how strongly the system is driven to consensus ---or to
maximum dissent, for the case of a negative temperature. We shall demonstrate this model by 
conducting several numerical experiments where a system with initially $N$ different ideas will
be subjected to temperature cycles, taking inspiration from the simulated-annealing algorithm, 
and the approach to global solutions will be characterized.

The rest of the paper is organized as follows. In Sec.~\ref{sec:CI}, we describe the principles of the CI experiment and the motivation for the proposed model. In Sec.~\ref{sec:model}, we introduce the model and a Markov process based on it that will define our dynamics. Numerical simulations are presented in Sec.~\ref{sec:sim} and our concluding remarks are summarized in Sec.~\ref{sec:conclusions}.

\section{The Collective Intelligence paradigm and semantic vectors}\label{sec:CI}
A CI experiment~\citep{garcia-egea:24} consists of a setup where participants are presented with a context describing a specific situation and then asked a series of questions, which they are required to answer. The participants are arranged on a square lattice with periodic boundary conditions, allowing them to view the responses of their four nearest neighbors. Consequently, they can either copy a neighbor's response or formulate a new one. 

To facilitate convergence, three dynamical mechanisms are implemented. The first is a copying mechanism, where a participant adopts a neighbor's response. The second is a permutation mechanism, in which a participant's position within the lattice is changed, exposing them to new neighbors and responses. The third is an extinction mechanism, where the least frequent responses are eliminated, leaving the response empty for those participants.

Experiments are organized into successive stages with evolving behaviors. In the first stage, participants operate in isolation to generate original responses. Starting from the second stage, the lattice structure becomes active, enabling participants to view their neighbors' answers. Subsequent stages incorporate the dynamical mechanisms designed to enhance collective interaction.

Regarding the copying dynamics, the transition probabilities can be determined by response frequencies. Let $f(n,t)$ be the frequency of the response held by node $n$ at time $t$. If a neighbor $n+\mu$ holds a response with frequency $f(n+\mu,t)$, the probability that user $n$ overwrites their response with that of the neighbor is given by:

\begin{equation}
P(n,n+\mu,t)= \frac {f(n+\mu,t)} {f(n+\mu,t)+f(n,t)}.
\label{eq:prob_Over}
\end{equation}

While this algorithm successfully generated consensus, it suffers from a fundamental physical limitation: nonlocality. A response change at a single node $n$ instantaneously alters the frequency count $f(n,t)$, thereby modifying the probabilities of every other node holding that response, regardless of their distance in the lattice.

To implement a dynamical system based on ideas, we must first define a quantitative metric for ``semantic similarity''. A direct approach would involve querying a generative LLM to rate the similarity between two texts on a continuous scale $[0,1]$. While this method offers high semantic accuracy, it presents an insurmountable computational bottleneck. In a typical CI experiment involving 1000 users, 5-10 new ideas are generated per second. If every new idea requires a call to the Application Programming Interface (API), the latency induced by thousands of API calls would render this solution unfeasible and seriously expensive.

A computationally efficient alternative is to map each textual response to a semantic vector $\phi$ using a pretrained embedding model. These models typically generate high-dimensional vectors, with a number $N$ of components from $N=256$ to 1024. Consequently, the semantic similarity between two responses $i$ and $j$ can be quantified by the scalar product between their semantic vectors $\phi(i)\cdot \phi(j)$. Unlike calling an LLM API, calculating the scalar product is negligible on modern GPUs and even the production of the vector can be done locally,
as large BERT models really involve only around 0.5 billions of parameters, easy to fit in the GPU vRAM.

As we demonstrate in this work, it is possible to define a local interaction based on semantic similarity that can similarly drive the system toward consensus while strictly adhering to locality principles. Furthermore, we will discuss how the additional dynamics defined in previous studies naturally emerge as implicit features of this new model.

\section{The model}\label{sec:model}

While it is possible to define a heuristic copying mechanism similar to the dynamics in Eq.~\eqref{eq:prob_Over} without an underlying statistical-mechanics framework, establishing a rigorous physical model offers significant advantages. It provides a solid foundation for understanding the system's behavior, analyzing its limiting cases, and systematically controlling parameters. Consequently, we propose a Hamiltonian-based model that not only satisfies the requirements of the previous section but also facilitates a simpler and more robust analysis of the experiments.

Let us consider a system where each participant $n$ holds a single response $i_n$. Each response is mapped to a normalized semantic vector $\phi(i_n)$, such that $\|\phi\| = 1$. Participants are placed on a square lattice of size $V=L\times L$ with periodic boundary conditions and nearest-neighbor interactions, so each node only {\it sees} its 4 adjacent neighbors $n+\mu$ and therefore interacts only with them.

We can define an energy for this system, which will be lowest when the alignment between semantic vectors is maximum (consensus) and increase when dissenting responses
coexist. The associated model can be defined as
\begin{equation}
H=-\sum_{n,\mu}\phi(i_n)\cdot \phi(i_{n+\mu}).
\label{eq:Action}
\end{equation}

This formulation \changes{is inspired by the} standard $O(N)$ model in $d=2$ dimensions with nearest-neighbor interactions; see, \emph{e.g.}, \cite{cardy:96,zinn-justin:05}. Here, $N$ represents the dimensionality of the semantic space, which typically ranges from $256$ to $1024$ in our case of use. Now, according to the Hohenberg-Mermin-Wagner theorem, continuous symmetries cannot be spontaneously broken with nearest-neighbors interactions in $d=2$~\cite{mermin:66,hohenberg:67}. This would suggest that consensus cannot be reached in such a model at any finite temperature. The same situation arises in flocking systems~\cite{vansaarlos:24}, where the problem is skirted due to the nonequilibrium nature of the models. Similarly, our dynamical law will not \changes{lead to} the equilibrium canonical distribution \changes{of the $O(N)$ model} \footnote{Long-range interactions are the other way around the Hohenberg-Mermin-Wagner theorem~\cite{halperin:19}. This is the case, for instance, of thermalized elastic membranes, where a flat phase is possible because 
in-plane phonons couple with height fluctuations, causing the bending rigidity to be renormalized~\cite{nelson:87}. In this work, however, we are interested in local interactions
only.}.

\changes {The first difference is that our set of available vectors does not span the entire continuous $O(N)$ hypersphere. Instead, the available state space, which we denote as $R(N)$, is restricted to the subset of active responses currently held by at least one user. Furthermore, this set is dynamic and nonincreasing under our Markov process: if a response is held by a single user who then copies a neighbor, that unique vector is permanently erased from the system, causing $R(N)$ to shrink. In a real experiment, $R(N)$ changes dynamically due to both this copying mechanism and the users' active modifications. } 

We propose the following Markov process:
\begin{enumerate}
\item{We choose a user $n$ in the network. The choice can be sequential or random.}
\item{ For a given node $n$, the set of new allowed states is restricted to the current responses possessed by its four nearest neighbors $\{i_{n+\mu}\}$, plus its current state. If the node retains its response, the change in energy is $\Delta H = 0$. If the node copies the response $i_{n+\nu}$ from a specific neighbor $\nu$, the change in energy is given by
\begin{equation}
    \Delta H= -\left[\sum_\mu \phi(i_{n+\nu})\cdot\phi(i_{n+\mu}) - \sum_\mu \phi(i_{n})\cdot\phi(i_{n+\mu})\right]
\end{equation}
}
\item Use $\Delta H$ to decide whether to accept the new state. Several choices for the transition matrix are possible, such as the standard Metropolis rule. We employ the heat-bath algorithm due to its efficiency in handling small discrete state spaces. The probability of choosing
response $i_{n+\nu}$ is, then,
\begin{equation}\label{eq:HB}
P(i_n\to i_{n+\nu}) = \frac{\exp{\left(-\beta \Delta H(i_n\to i_{n+\nu})\right)}}{\sum_\mu \exp{\left(-\beta \Delta H(i_n \to i_{n+\mu})\right)}}\, .
\end{equation}
In this equation, the sum includes the case $\mu=0$, corresponding to no change. \changes{Our dynamical law includes a variable parameter $\beta$, which, by analogy with equilibrium statistical mechanics, we shall refer to as an (inverse) temperature.}
\end{enumerate}

\changes{
It is important to distinguish this process from the canonical dynamics of the $O(N)$ model, where a tentative change can be to any $O(N)$ vector. First, such a move would, in almost all cases, explore regions of the semantic-vector space that do not correspond to any sensible human response. Second, because vectors in $R(N)$ can permanently disappear under our dynamics, the system naturally tends toward absorbing states. Therefore, unlike an ergodic transition matrix in the standard $O(N)$ model, the dynamics proposed above are capable of breaking the continuous symmetry and, hence, achieving consensus. Hence, our $\beta$ will not represent 
true thermal equilibrium but, as we shall explain later, it will still control 
energy fluctuations and allow us to define an ``annealing'' procedure to drive
the system towards ordered states.}

To study the evolution of this model, we define two metrics. The first is the {\it living-response} count $l_\text{v}$, defined as the number of different responses on the lattice, \changes{therefore, $l_\text{v}=|R(N)|$}. Note that under our proposed dynamic, this metric can only decrease. If the last user holding response $i_n$ copies a neighbor, $i_n$ is permanently extinguished from the system. The second is the {\it semantic energy} $e_\text{s}$ defined as
\begin{equation}
e_\text{s}=\dfrac{H}{4V}=-\frac{1}{4V} \sum_{n,\mu}\phi(n)\phi(n+\mu) 
\label{eq:Esem}
\end{equation}

Note that this is an intensive metric, $e_\text{s}\in[-1,1]$, where $-1$ corresponds to perfect global alignment. Furthermore, we can decompose $e_\text{s}$ into the {\it local semantic energy} $e_\text{ls}(n)$ of a node $n$ as
\begin{equation}
e_\text{ls}(n)=-\frac{1}{4} \sum_{\mu}\phi(n)\phi(n+\mu) 
\label{eq:Elocalsem}
\end{equation}
ensuring that $e_\text{s} = \frac{1}{V}\sum_n e_\text{ls}(n)$.

\section{Simulations}\label{sec:sim}

To study the model, we perform Monte Carlo simulations on lattices with periodic boundary conditions. Let $V=L\times L$ be the size of the lattice, which we initialize with $V$ different responses, one for each node. Subsequently, we construct the similarity matrix $s(i,j)$ where each element $(i,j)$ represents the scalar product between responses $i$ and $j$, $s(i,j) = \phi(i)\phi(j)$.

\begin{figure}[ht]
    \includegraphics[width=1.0\linewidth]{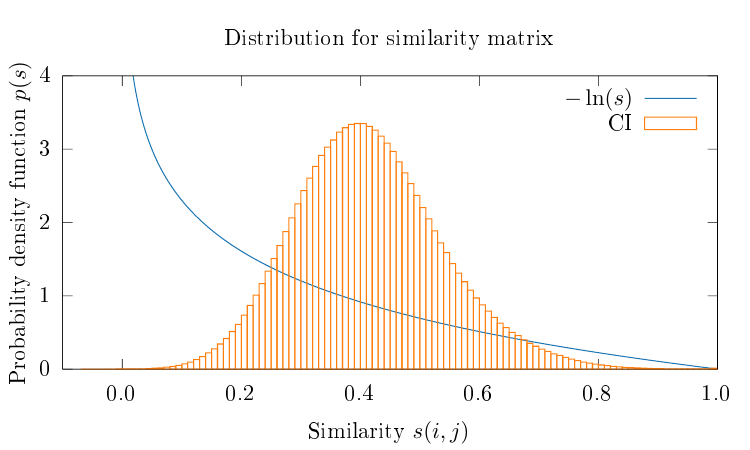}
    \caption{Distributions used in the experiments. The graph of the function $-\ln(s)$ represents the probability density function of $P(X \cdot Y)$ where $X$ and $Y$ are independent random variables with a uniform distribution on $[0,1]$. The CI is the histogram of the distribution of the similarity matrix between the embeddings of the responses from a real CI experiment with more than 5000 different responses.}
    \label{fig:distribution}
\end{figure}

We have studied the behavior of the model for two different starting conditions. We first consider a synthetic dataset, where the similarity matrix $s(i,j)$ is generated 
from the expected distribution $P(X\cdot Y)$ of two independent random variables $X$ and $Y$ with a uniform distribution on $[0,1]$. The density probability function for a given similarity $s$ is $p(s)=-\ln(s)$. \changes{We initialize the system such that every user starts with a unique response.}

Alternatively, we use data from a real CI experiment where the users generated more than 5000 different responses (see Appendix~\ref{app:CI experiment}). In this case, we computed the semantic vectors using the Arctic Embed 2.0 model, specifically \texttt{snowflake-arctic-embed-l-v2.0}~\citep{snowflake:24} with an output dimension of 256. The model weights can be obtained from the Hugging Face repository in~\cite{snowflakerep}. \changes{For each simulation, we initialize the lattice by randomly sampling a unique subset of $V$ responses from the total pool, ensuring that every user starts with a distinct response.} %

\changes{In all simulations, the lattice size is set to $V=32\times32=1024$.}

\changes{As can be observed in \autoref{fig:distribution}, the scalar products $s(i,j)$ for both the synthetic dataset and the real CI experiment take only positive values. For the real CI data, this strict positivity is a direct consequence of the embedding model's training. Because the similarity between any pair of interacting users is always greater than zero, the semantic energy $e_s$ of our system is naturally restricted to negative values ($e_s < 0$). Thus, while the theoretical bounds defined in Eq. \eqref{eq:Esem} allow for $e_s \in [-1, 1]$, the dynamics in our specific simulations will exclusively take place in the negative-energy domain.}

To study the evolution of the system, we will measure the semantic energy $e_\text{s}$, Eq.~\eqref{eq:Esem}, and the number of living responses $l_\text{v}$.

Taking inspiration from the Simulated Annealing algorithm, 
we consider several possibilities for temperature-cycling protocols, aiming to study consensus and dissent phenomena and to accelerate the system's convergence. These three options are:
\begin{itemize}
    \item {\bf Standard}: We start with $\beta$ close to 0 and increase it during the Monte Carlo process to positive values. In our case, we start with 1 and end with 8.
    \item {\bf Negative Standard}: We start with a negative $\beta$ close to 0 and decrease during the Monte Carlo process to negative values. In our case, we start with $-1$ and end with $-8$.
    \item \changes{{\bf Alternating positive and negative $\beta$}:  We start with negative values of $\beta$ and increase them to positive values of $\beta$. In our case, we start with $-8$ and end with 8. In this way we have a consensus-dissent procedure, where users are surrounded alternatively with similar or differing ideas.}
\end{itemize}

\changes{For each of these options, the annealing schedule is structured as a series of thermal cycles. The range of $\beta$ is divided into 50 discrete steps to form a single cycle. This cycle is repeated 10 times, resulting in a total of $N_{iter}=500$ steps for $\beta$ following a sawtooth pattern. Furthermore, at each discrete step of $\beta$, the system is relaxed by performing 10 full Monte Carlo sweeps. Measurements are taken only after this relaxation of the system.}

\begin{figure}[t]
    \centering
    \includegraphics[width=1.0\columnwidth]{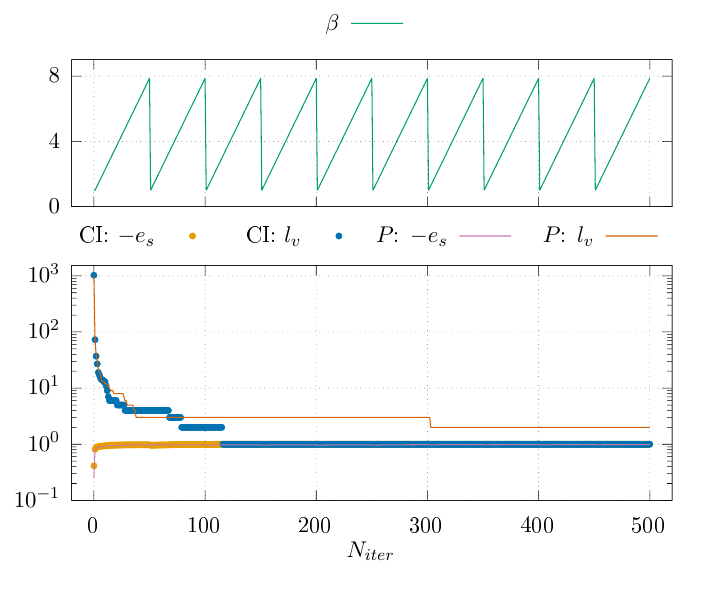}
    \caption{Example of evolution for the Standard annealing process with $\beta\in[1,8]$. Results for the CI distribution are represented with points, and results for the $P(X \cdot Y)$ distribution are represented with lines.}
    \label{fig:pos_evol}
\end{figure}

\autoref{fig:pos_evol} shows a typical evolution with the Standard annealing process. This option leads the system to a configuration where all response vectors are aligned, which we call ferromagnetic, using the standard language for the $O(N)$ model. The system starts with $l_ \text{v} = V$ and $e_ \text{s}$ in a maximum. As the system evolves, $l_ \text{v}$ decreases and the energy $e_ \text{s}$ tends to $-1$. Eventually, the system reaches equilibrium where only one response is left, $l_ \text{v}=1$ and $e_ \text{s} = -1$, as clearly seen for the CI distribution. We can say that this is a state of consensus, because all nodes in the lattice are surrounded by neighbors with the same response, that is, people that share the same idea.

\begin{figure}[t]
    \centering
    \includegraphics[width=1.0\linewidth]{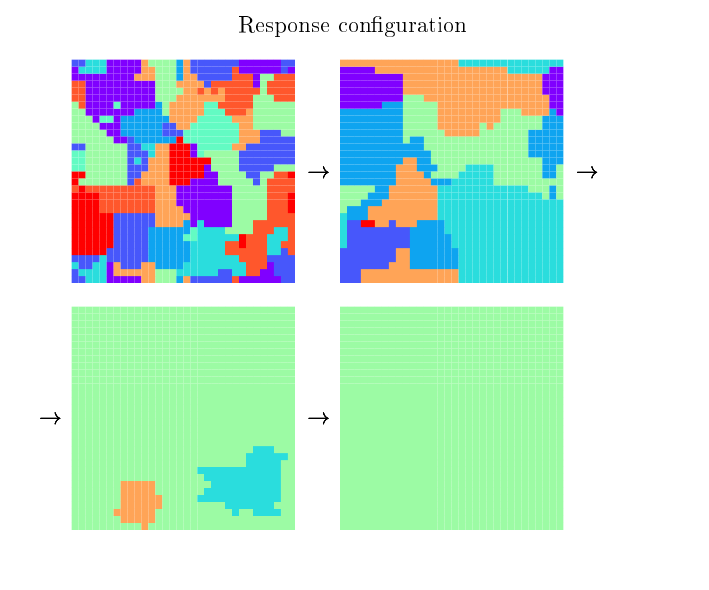}
    \caption{Snapshots of the response configuration in the ferromagnetic regime. Nodes are colored according to the response (idea) they represent. Arrows indicate the direction of time evolution. The corresponding local energy distribution for each frame is shown in \autoref{fig:ferro_energy_conf}.}
    \label{fig:ferro_conf}
\end{figure}

\begin{figure}[ht]
    \centering
    \includegraphics[width=1.0\linewidth]{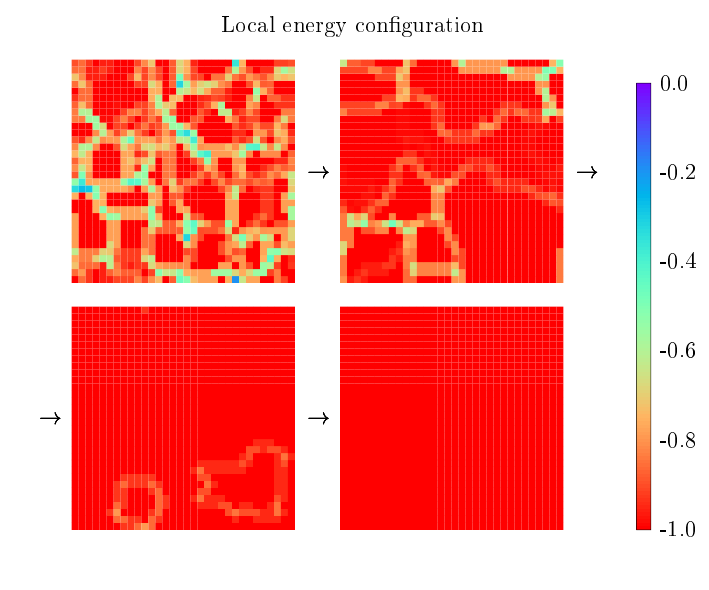}
    \caption{Evolution of the local energy configuration in the ferromagnetic state. Nodes are colored according to their local energy, as defined in \eqref{eq:Elocalsem}. Each panel corresponds to the respective time step shown in \autoref{fig:ferro_conf}.}
    \label{fig:ferro_energy_conf}
\end{figure}

\begin{figure}[t]
    \centering
    \includegraphics[width=1.0\columnwidth]{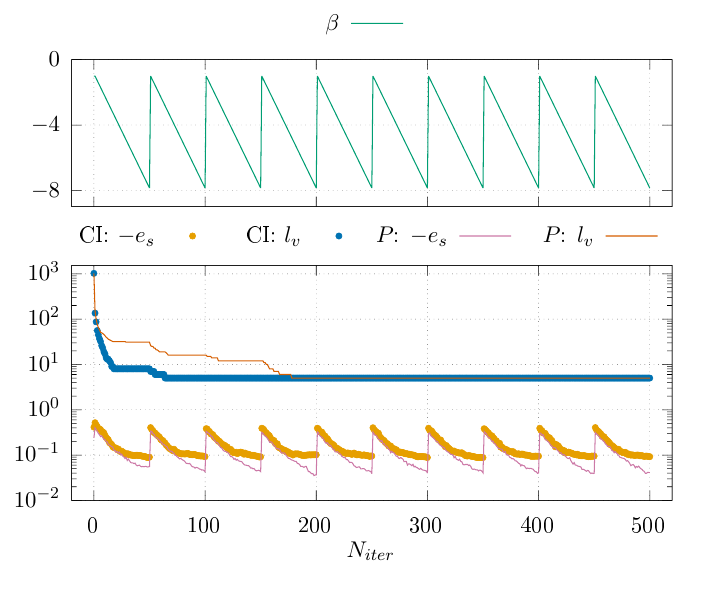}
    \caption{Example of evolution for the Negative Standard annealing process  with $\beta\in[-8,-1]$. Results for the CI distribution are represented with points, and results for the $P(X \cdot Y)$ distribution are represented with lines.}
    \label{fig:neg_evol}
\end{figure}

We illustrate the evolution of the system on the lattice in \autoref{fig:ferro_conf}, which displays four snapshots of the response configuration. Initially, the system exhibits a disordered multi-domain structure characterized by numerous small clusters. As time progresses, a coarsening process takes place where smaller clusters vanish while larger domains expand, eventually leading to a state where a single response dominates (consensus). In \autoref{fig:ferro_energy_conf}, we depict the corresponding evolution of the local energy. In the bulk of the domains, the local energy assumes its minimum value, $e_ \text{s}=-1$. Higher energy values are strictly confined to the cluster boundaries, highlighting the energy cost associated with lack of local consensus.

Using the Negative Standard annealing process, we can see a different behavior that leads to an antiferromagnetic configuration. Again, the system starts with $l_ \text{v} = V$, but \autoref{fig:neg_evol} shows a different behavior for the energy $e_ \text{s}$. As $\beta$ decreases, the energy also decreases. Note that $e_ \text{s}$ never approaches $-1$ unlike in the previous case, even though $l_ \text{v}$ decreases. The equilibrium in that case is $l_ \text{v}=2$ where both responses are arranged in the lattice in a checkerboard pattern. This represents a state of dissent because all nodes in the lattice are surrounded by neighbors with a different response, that is, people that do not share the same idea.

\begin{figure}[t]
    \centering
    \includegraphics[width=1.0\linewidth]{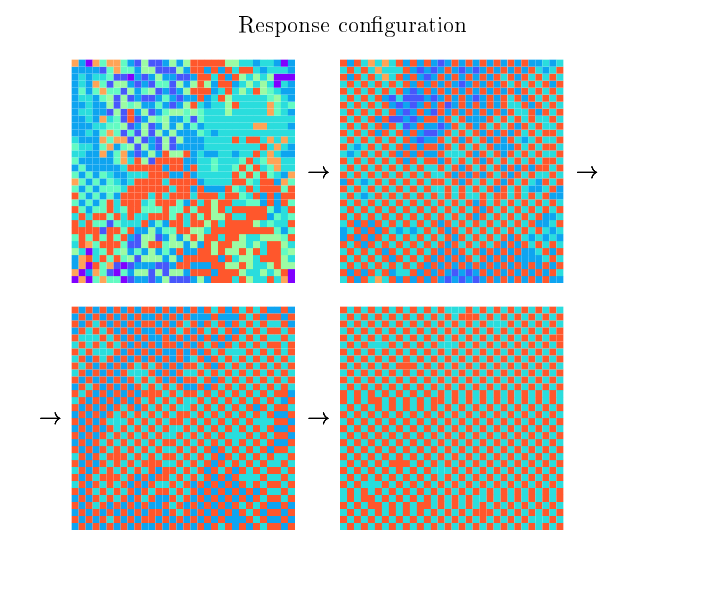}
    \caption{Snapshots of the spatial configuration in the antiferromagnetic regime. Nodes are colored according to the response they represent. Unlike the ferromagnetic case, the system evolves into a checkerboard-like pattern. Arrows indicate the direction of time evolution. The corresponding local energy distribution is shown in \autoref{fig:af_energy_conf}.}
    \label{fig:af_conf}
\end{figure}

\begin{figure}[t]
    \centering
    \includegraphics[width=1.0\linewidth]{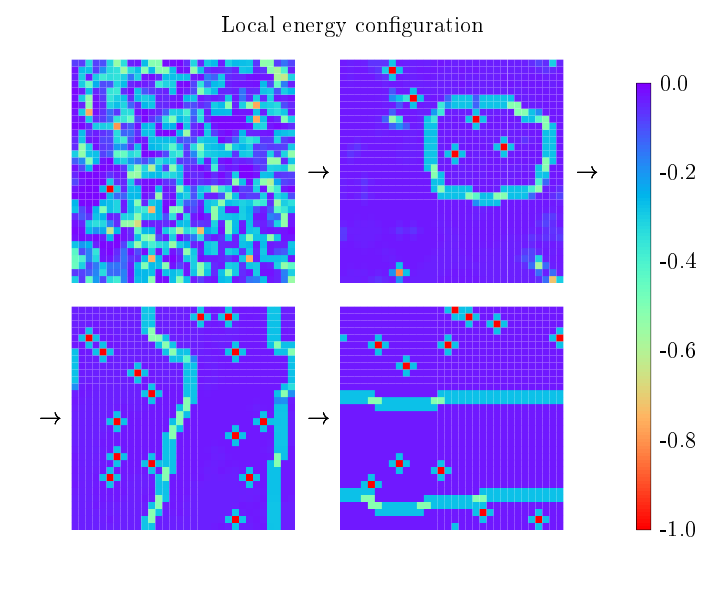}
    \caption{Evolution of the local energy landscape in the antiferromagnetic state. Nodes are colored according to their local energy defined in \eqref{eq:Elocalsem}. In this regime, the system stabilizes at higher energy values, with lower energy excitations confined to domain boundaries. Each panel corresponds to the frames in \autoref{fig:af_conf}.}
    \label{fig:af_energy_conf}
\end{figure}

The evolution of the system in the antiferromagnetic regime is illustrated in \autoref{fig:af_conf}. Initially, the system displays a highly disordered configuration with a diverse set of responses. However, in contrast to the ferromagnetic phase, the evolution does not lead to the formation of monochromatic clusters. Instead, the responses arrange themselves into a checkerboard-like structure, where each node maximizes its semantic difference relative to its nearest neighbors.

This behavior is quantitatively captured by the local energy configuration shown in \autoref{fig:af_energy_conf}. As the system evolves, the local energy in the bulk of the domains stabilizes at a relatively high value close to zero. This high-energy state indicates that neighboring vectors are nearly orthogonal to each other. Interestingly, the energy only drops to lower values at the boundaries where the perfect checkerboard pattern is disrupted. This confirms that, even though the checkerboard pattern comprises distinct responses, the system successfully maintains a state of global diversity by minimizing local semantic overlap (dissent).

\begin{figure}[t]
    \centering
    \includegraphics[width=1.0\columnwidth]{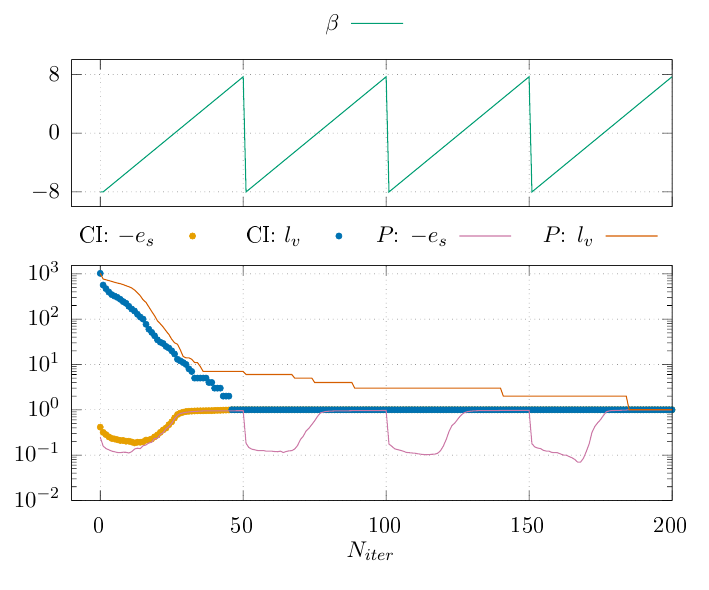}
    \caption{Example of evolution for the Consensus-dissent annealing process with $\beta\in[-8,8]$. Results for the CI distribution are represented with points, and results for the $P(X \cdot Y)$ distribution are represented with lines.}
    \label{fig:sierra_evol}
\end{figure}

When using the Consensus-dissent strategy, we find that  convergence is achieved sooner. \autoref{fig:sierra_evol} shows a typical example of the evolution for this Consensus-dissent annealing process. Here $\beta$ starts with negative values, where the energy is in a local minimum and when $\beta$ increases to positive values, the energy rapidly approaches $-1$. 

In all the experiments, the initial values of $l_ \text{v}$ and $e_ \text{s}$ are what we expect. $l_ \text{v}=V$ is satisfied because we start with a unique response for each node. 
For $e_ \text{s}$, at the beginning, the expected value is
\begin{equation}
    e_ \text{s}^{t=0}=-\dfrac{1}{4V}\sum_{n,\mu}E[\phi(n)\cdot \phi(n+\mu)].
\end{equation}
Where $\phi(n)\cdot \phi(n+\mu)$ at the beginning have the distribution shown in \autoref{fig:distribution}, therefore, we can write $E[\phi(n)\cdot \phi(n+\mu)]=E[s]$. 

In the case of $P(X\cdot Y)$ where $p_S(s)=-\ln (s)$, the expected value is
\begin{equation}
    E[s] = -\int_0^1 s \ln(s) \text{d}s = \dfrac{1}{4}
\end{equation}
Therefore,
\begin{equation}
    e_\text{s}^{t=0}=-\dfrac{1}{4V}\sum_{n,\mu}\dfrac{1}{4}=-\dfrac{1}{4}.
\end{equation}

In the case of CI experiment, $E[s]$ is given by the mean of the histogram in \autoref{fig:distribution}, which is $E[s]\approx 0.4$. Therefore 
\begin{equation}
    e_\text{s}^{t=0}\approx-\dfrac{1}{4V}\sum_{n,\mu}0.4=-0.4.
\end{equation}

In all cases, the evolution starts at the expected initial values.

Regarding the initial distribution, the only change in the system behavior is that convergence is achieved faster with the CI distribution compared to $P(X\cdot Y)$. On the other hand, the distributions used have the property that, due to the postraining of the model, the similarity is always positive, therefore in our system $e_\text{s} < 0$.

\section{Conclusions}\label{sec:conclusions}

We presented a semantic-vector model for CI experiments.
Textual responses from participants are converted into N-component semantic vectors ($N=256$ to $1024$) with the use of an embedding model. 
The Hamiltonian of the standard $d=2$ $O(N)$ model, based on the scalar product between nearest-neighbor vectors on a square lattice, is then employed to define an energy that quantifies the level of consensus in the system. This energy function can then be used to define a dynamical process based only on
local interactions. This approach both simplifies the method used in previous work~\cite{garcia-egea:24}, which relied on nonlocal counting of response frequency, and provides a more faithful description of the state of the system, since it can account for semantically similar but not textually identical responses.

With this new model, we can overwrite responses using only local information and we can extinguish responses. Furthermore, we can adjust the value of a coupling constant $\beta$ so each node naturally becomes surrounded by similar ideas ($\beta > 0$) or distinct ideas ($\beta < 0$). Previously, dissent in the system could only be explored by exchanging the lattice sites of participants in the CI experiment~\cite{garcia-egea:24}.

We show that with this model we can generate consensus for $\beta > 0$, where all nodes in the lattice share the same response, and we can generate maximum dissent for $\beta < 0$, where responses are placed in a checkerboard pattern.

\begin{acknowledgments}
This work was partially supported by Ministerio de Ciencia, Innovación y Universidades (Spain) and by the European Regional Development Fund (MCIU/AEI/10.13039/501100011033/FEDER, UE) through grants No. PID2024-158623NB-C22 and No. PID2022-136374NB-C22. We were also partly funded by Gobierno de Aragón (research group E30\_23R).
\end{acknowledgments}

\appendix

\section{Collective Intelligence experiment}\label{app:CI experiment}
We used responses generated from a Collective Intelligence experiment involving 656 adolescent participants \citep{orejudo:22}. The scenario presented involved two youths who experienced an incident of grooming. Six questions were posed regarding this context. In total, more than 5000 unique responses were generated in the experiment.

For each simulation, we take $V=L\times L$ different responses and compute their respective embeddings with the Arctic Embed 2.0 model~\citep{snowflake:24}. Each response and embedding is randomly placed on the lattice to start the simulation.

\end{document}